# FACTORS INFLUENCING SAUDI CUSTOMERS' DECISIONS TO PURCHASE FROM ONLINE RETAILERS IN SAUDI ARABIA: A QUANTITATIVE ANALYSIS

Rayed AlGhamdi, Ann Nguyen, Jeremy Nguyen and Steve Drew
*School of ICT, Griffith University*

**ABSTRACT**

This paper presents the preliminary findings of a study researching the diffusion and the adoption of online retailing in Saudi Arabia. It reports new research that identifies and explores the key issues that positively and negatively influence the decision of Saudi customers to buy from online retailers in Saudi Arabia. Although Saudi Arabia has the largest and fastest growth of ICT marketplaces in the Arab region, e-commerce activities are not progressing at the same speed. While the overall research project involves exploratory research using mixed methods, the focus of this paper is on a quantitative analysis of responses obtained from a survey of Saudi customers, with the design of the questionnaire instrument being based on the findings of a qualitative analysis reported in a previous paper. The main findings of the current analysis include a list of key factors that affect Saudi customers' purchase from Saudi online retailers, and quantitative indications of the relative strengths of the various relationships.

## 1. INTRODUCTION

The revolution of electronic commerce (e-commerce) has started in the 90s in the developed world. Many commercial organizations around the world have introduced e-commerce models in their businesses, seeking the many benefits that the online channel can provide (Laudon and Traver 2007). Basically, e-commerce is commerce enabled by Internet technologies, including pre-sale and post-sale activities (Whiteley 2000; Chaffey 2004) and online retailing is a model of business to customer (B2C) e-commence which is online version of traditional retail (To & Ngai 2006). Since 2000, e-commerce's rapid growth is obvious in the developed world. Global e-commerce spending has currently reached US$10 trillion and was US$0.27 trillion in 2000 (Kamaruzaman, Handrich & Sullivan 2010). The United States, followed by Europe, constitutes the largest share with about 79% of the global e-commerce revenue (Kamaruzaman, Handrich & Sullivan 2010). However, the African and Middle East regions have the smallest share with about 3% of the global e-commerce revenue (Kamaruzaman, Handrich & Sullivan 2010).

Regarding Saudi Arabia, the world's largest oil producer (CIA 2009), e-commerce is still underdeveloped. Although Saudi Arabia has the largest and fastest growth of ICT marketplaces in the Arab region (Saudi Ministry of Commerce 2001; Alotaibi and Alzahrani 2003; U.S. Commercial Services 2008; Alfuraih 2008), e-commerce activities are not progressing at the same speed (Albadr 2003; Aladwani 2003; CITC 2007). Only 9% of Saudi commercial organizations, mostly medium and large companies from the manufacturing sector, are involved in e-commerce (CITC 2007). This paper is part of a research project studying the diffusion of online retailing in Saudi Arabia. The focus of this paper is on the investigation of factors that affect Saudi customers' decisions to purchase from online retailers in Saudi Arabia.

## 2. LITERATURE REVIEW

Information and Communication Technology (ICT) plays a significant role in the countries' economies. Over the last decade, the Saudi government has concentrated on this field to become the largest and fastest growing ICT marketplace in the Arab region (Saudi Ministry of Commerce 2001; Alotaibi and Alzahrani 2003; U.S. Commercial Services 2008; Alfuraih 2008). The Saudi government has introduced policies to encourage public and private organizations to adopt ICTs (Al-Tawil, Sait and Hussain 2003). In Saudi Arabia, so far, the effort towards e-commerce development have not reached its originally stated aspirations; neither what it sees as the world's expectations of a country of the level of importance and weight in the global economy like Saudi Arabia. The government support for e-commerce development seems to be missing which represents a key for online retail growth in Saudi Arabia (AlGhamdi & Drew 2011and AlGhamdi, Drew & Alkhalaf, 2011). Official/government information/documents about e-commerce in Saudi Arabia are not sophisticated. The official discussion to introduce e-commerce in Saudi Arabia started in 2001. In that year, Saudi Ministry of Commerce established a permanent technical committee for e-commerce including members from the Ministries of Commerce, Communication and Information Technology and Finance. It also includes members from the Saudi Arabian Monetary Authority (SAMA) and King Abdulaziz City for Science and Technology (KACST) (Saudi Ministry of Commerce 2001). The roles of this committee are to follow the developments in the field of e-commerce and take the necessary steps to keep pace with them. However, this committee does no longer exist. The role of e-commerce supervision in the country has been transferred to the Ministry of Communications and Information Technology since 2005. So far, the efforts of e-commerce support by Ministry of ICT are hapless. A phone call made to Ministry of ICT followed by e-mail on April 2011 seeking further information about e-commerce support and development. The answer was that the Ministry of ICT in Saudi Arabia is still in its early stages of studying e-commerce. Currently, they are conducting a survey on e-commerce in Saudi Arabia and a report may be published in May/June 2011.

According to EUI (2010) report which assessed the quality of 70 countries' ICT infrastructure and the ability of their government, businesses and people to use ICT, Saudi Arabia ranked 52 in e-readiness. The extent of Internet access in Saudi Arabia indicates its e-commerce readiness (Sait, Altawil, and Hussain 2004). The Internet was introduced in Saudi Arabia in 1997 (Alzoman 2002). Only King Abdulaziz City for Science and Technology (KACST) provides Internet access; therefore, all Internet users in Saudi Arabia go through KACST (Algedhi 2002; Saudi Internet 2007b). The Internet users increased from one million (5% of the population) in 2001 to 11.2 million (41%) in 2010 (MCIT 2010). "Broadband subscriptions have grown from 64,000 in 2005 to over 3.2 million at the end of Q3 2010" (MCIT 2010). However, broadband subscriptions remain very low compared to the developed nations.

The Arab Advisor Group carried out an extensive survey in mid-2006, targeting Internet users in four Arab countries (Saudi Arabia, UAE, Kuwait and Lebanon). The survey covered Internet usage and, e-commerce activities in these countries. While UAE ranked first in the rate of annual spending on e-commerce per capita, Saudi Arabia ranked first in the overall money spent on e-commerce activities. As for the prevalence of e-commerce activities among the population, UAE ranked first at 25.1%, Saudi Arabia second at 14.3%, Kuwait third at 10.7% and Lebanon last at 1.6% (AAG 2008). A recent survey of Saudi Arabia's Internet users found that around 3.1 million Saudis have purchased online. Airline tickets and hotels bookings take the largest percentage of these purchases (ACG 2009, AAG 2011). Although the youth are the majority population of the six Gulf countries (GCC: KSA, UAE, Kuwait, Oman, Qatar and Bahrain), increasingly using the latest technologies, online shopping remains under-developed, mainly because of "the relatively low levels of internet usage and low credit card penetration" (ACG 2009). Approximately 45% of GCC populations have purchased online.

Several studies have been conducted to discover the reasons behind the slow e-commerce developments in the Arab world in general and Saudi Arabia in particular. The reasons were mainly involved ICT infrastructure, trust and privacy issues, cultural issues, and the absence of clear regulations, legislation, rules and procedures on how to protect the rights of all involved parties (Albadr 2003; Aladwani 2003; Al-Solbi and Mayhew 2005; CITC 2006; Alfuraih 2008; Alraw; Sabry 2009 and Alghaith, Sanzogni and Sandhu, 2010). Although Saudi Arabia contributes to the efforts of UNCITRAL (United Nations Commission into International Trade Laws) (Saudi Ministry of Commerce 2001), there is a need to have major development in terms of e-commerce regulations, legislations and rules to protect the rights of all parties involved in e-commerce transactions (Albadr 2003; Al-Solbi and Mayhew 2005).

Other challenges involve the mailing system (Alfuraih 2008). Before Saudi Post was established in 2005, individuals had no home addresses (Saudi Post 2008); therefore, to receive mail, individuals had to subscribe to have a mailbox in the post office (Alfuraih 2008). In 2005, the postal delivery to homes and buildings was approved by Saudi Post (Alfuraih 2008; Saudi Post 2008). While this service is still relatively new, Saudi Arabia is very late in providing individual addresses. Problems with adopting this service might be the citizens' lack of awareness of this service or the importance of mailboxes, their ignorance of the direct addresses for their houses with numbers and streets names, or their mistrust of receiving their mail in this way. Consequently, more efforts are needed to motivate the citizens owning house mailboxes and solve the problems that they face.

At the end of 2010, Saudi Post launched an electronic mall, "the first online marketplace in Arabic and English" (E-mall 2010), giving Saudi retailers the chance to sell their products online and benefit from cheap delivery fees. In March 2011, e-mall administrators revealed that to date, there are 50 sellers, 50,000 buyers, which is a 10% increase in buyers and sellers. 2,000 deals have taken place, totalling 2 million KSR, with the preferred payment method being SADAD (Al-Mohamed 2011). It seems Saudi Post adopting online mall to encourage more citizens subscribe in their services including having a home mailbox.

The culture of people to buy in Saudi Arabia is still a key factor influences retailers to adopt online sale channel. The difficulty to attract customers buying online is the first answer that you get when you ask a retail decision maker in Saudi Arabia why you do not adopt and use online retail channel. There is an emphasis on this factor making it a key concept for deterring the diffusion of e-retail systems in Saudi Arabia (AlGhamdi, Drew & Al-Ghaith 2011). However, the spending of online retailing in Saudi Arabia is growing. Online retail sector size estimated to about SAR 3 billion (US$1= SAR 3.75). This figure represents 20% of the total Electronic Trading in Saudi Arabia. The average value of what a customer pays for each online purchase is about SAR400 (Hamid 2011). This spending of growing community of online customers vs the slowness of retailers to introduce online sale channel in Saudi Arabia is an indicator that retailers in Saudi Arabia are not realizing the importance of online retail yet.

Up to date, no single study has been founded studding customers behaviours, drivers and inhibitors to purchase online from retailers in Saudi Arabia. As stated by retailers, the major concerns for them that the customers' behaviour to buy online is frustrated and that is what discourages them introducing online sales channel (AlGhamdi, Drew & Al-Ghaith 2011). In contrast, Internet customers in Saudi Arabia are growing and a number of their online spending goes overseas. For this reason, this research explores the factors influencing Saudi customers' decisions to purchase from online retailers in Saudi Arabia. The study, firstly, used qualitative approach to explore the issues. The purpose of this paper is to follow up the qualitative study in order to test the findings in a wider sample.

The qualitative study by AlGhamdi, Drew & AlFaraj (2011) established a list of factors that inhibit customers to purchase online from an e-retailer in the KSA. These inhibitors are (1) lack of home mailbox, (2) feeling uncomfortable paying online with a credit card, (3) do not know e-retailers in Saudi Arabia, (4) lack of experience in buying online, (5) not having easy and fast access to the Internet, (6) lack of physical inspection of a product, (7) personal information (name, mobile number, e-mail etc) privacy, (8) lack of clear regulations and legislation for e-commerce in the KSA, (9) lack of the language understanding if the website or part of it is in English, (10) not trusting e-retailers in Saudi Arabia. The qualitative study also established a list of incentives that may encourage customers to purchase online from e-retailers in the KSA. These enablers are (1) competitive prices, (2) owning a home mailbox, (3) easy access and fast speed of the Internet, (4) provision of educational programs, (5) local banks make it easy to own a credit card, (6) professional and easy to understand design of the e-retailer's website, including showing complete specifications with photos of the products, (7) existence of a physical shop besides the online shop (Brick and click), (8) existence of online payment options other than credit cards, (9) existence of government support, supervision and control. So, the purpose of this paper is to establish numerically the relative strengths of these inhibiting and enabling factors.

## 3. RESEARCH METHODOLOGY

The whole project studying online retail in Saudi Arabia is built on the combination of qualitative and quantitative approaches. Qualitative study conducted first for exploration purpose and followed by quantitative approach based on qualitative findings for testing purpose. This type of approach is called exploratory mixed methods design (Creswell 2008, p. 561) which is done "to explore a phenomenon, and then [collect] quantitative data to explain relationships found in the qualitative results" (Creswell 2008, p. 561). The mixed methods approach helps to provide an in-depth investigation of the research problem (Morse 2003; Johnson & Onwuegbuzie 2004; Greene 2007; Alise and Teddlie 2010; Feilzer 2010).

While the qualitative study was conducted in a separate research (AlGhamdi, Drew & AlFaraj 2011), the focus of this paper is on quantitative analysis based on the qualitative findings. In the qualitative study (AlGhamdi, Drew & AlFaraj 2011), interviews were conducted with 16 Saudi participants (eight males and eight females) aged from 16 to 45 years. A qualitative content analysis was used to identify the factors that positively and negatively influence customers' decision to purchase from online retailers in Saudi Arabia.

In this paper, a questionnaire survey based on the qualitative study's findings is used to gain more information about the relative strengths of these factors (for the survey questionnaire, please contact the main author). Typically a question that asks for information about the participant's background and attributes would provide a set of choices, plus an open answer (e.g., "other") where the participant could insert additional information if he or she wishes. The two key questions are "What factors inhibit or discourage you from buying online from e-retailers in Saudi Arabia?" and "What would enable you to buy online from e-retailers in Saudi Arabia?" The participants would be given a list of 11 options to select from for the former question, and 10 options for the latter (in each case, the last option is "other reasons"). Respondents may select as many of the available options as they wish, including the open answer. The survey questions are in Arabic.

Two forms were distributed. Paper and online surveys were collected, with 50% for each form. The aim is to collect answers from 700 participants which are still ongoing; however, up-to-date the total number of participants reported in this paper reached 412. They were selected randomly with consideration to represent 50% of each gender, cover different age groups, and come from different cities in Saudi Arabia.

## 4. RESULTS AND DATA ANALYSIS

This section presents a summary and analysis of the responses collected to date from 412 participants. Male respondents account for 50% of the sample, and female respondents 50%. Respondents aged 15-25 represent 27.2% of the sample, compared with 49.0% for those aged 26-35 and 23.8% for those aged 36 and over. About 28.6% of the sample are residents of the capital city (Riyadh), compared with 26.0% for Jeddah, 11.7% for Al-Baha, 28.2% for residents of smaller cities, and the remaining 5.8% for residents of smaller urban centres.

Table 1 reports our findings with respect to the relative importance of factors that inhibit Saudi customers from making online purchases from Saudi e-retailers. In this table, the inhibitors are listed in the order in which they are presented to the respondents. Figure 1 illustrates the same information, but with each inhibitor being ranked according to its relative weight, from being most frequently selected to least.

Table 1. Inhibitors of online purchases by Saudi customers from Saudi vendors

| Identifier | Inhibitor | Selected by % of respondents | Rank |
|---|---|---|---|
| IN1 | Lack of experience in buying online | 40.8 | 4 |
| IN2 | Not trusting e-retailers in Saudi Arabia | 38.6 | 5 |
| IN3 | Not having easy and fast access to the Internet | 7.3 | 10 |
| IN4 | Cannot inspect product, worry about quality | 58.0 | 1 |
| IN5 | Lack of mailbox for home | 31.8 | 7 |
| IN6 | Do not know e-retailers in Saudi Arabia | 38.6 | 6 |
| IN7 | Not comfortable paying online using credit card | 27.7 | 8 |
| IN8 | Do not understand if website (or part) is in English | 20.9 | 9 |
| IN9 | No clear regulations & legislations for EC in KSA | 53.4 | 2 |
| IN10 | Don't trust that personal info. will remain private | 44.7 | 3 |
| IN11 | Others | 6.1 | 11 |

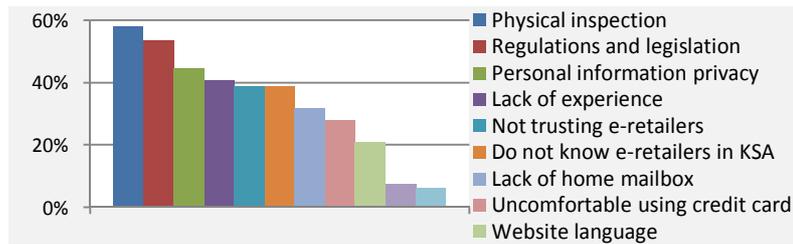

Figure 1. Factors inhibiting online purchases by Saudi customers from Saudi e-retailers

From Table 1 and Figure 1, it can be seen clearly that the most serious inhibitors tend to be related to a lack of trust and/or experience with online purchasing. For example, Inhibitor IN4 (Customers cannot inspect product and worry about product quality) is ranked 1, being cited by 58.0% of the respondents. Similarly, IN10 (Customers do not trust that their personal information, such as name, mobile phone number, etc, will remain private) is selected by 44.7% of respondents (ranked 3). IN2 (Customers do not trust e-retailers in Saudi Arabia) which refers directly to a lack of trust in Saudi e-retailers, is cited by 38.6% of respondents (rank 5). In the same vein, 40.8% of the respondents indicate that lack of experience in buying online (IN1) is a major inhibitor in their case (ranked 4), and 38.6% indicate that they don't know Saudi e-retailers well (IN6, ranked 6). It appears that in the minds of many Saudi customers, a lack of clear government regulations and legislations on e-commerce may have been a key contributor to this general lack of trust and experience: IN9 is chosen by 53.4% of respondents (ranked 2).

Interestingly, while a lack of a home mailbox (IN5, ranked 7) or being uncomfortable with using a credit card to make online payments (IN7, ranked 8) may deter significant percentages of the people in the sample (31.8% and 27.7%, respectively), numerically these are clearly less important than the trust/experience issues. It is rather re-assuring to find that lack of a command of English (IN8) is a relatively minor issue (20.9%, ranked 9). In view of recent IT infrastructure developments in Saudi Arabia it is not surprising that lack of ready access to fast Internet connections (IN3) is largely a non-issue (7.3%, ranked 10).

Table 2 and Figure 2 present basic survey results with regard to factors that would tend to enable or encourage customers to purchase online from Saudi e-retailers.

Table 2. Enablers of online purchases by Saudi customers from Saudi vendors

| Identifier | Enabler | Selected by % of respondents | Rank |
|---|---|---|---|
| EN1 | Owning house mailbox | 37 | 5 |
| EN2 | Easy access & fast Internet speed | 33 | 7 |
| EN3 | Competitive prices | 57 | 3 |
| EN4 | Local banks make owning credit cards easier | 21 | 9 |
| EN5 | Provision of educational programs | 29 | 8 |
| EN6 | Government support, supervision & control | 58 | 2 |
| EN7 | Well-designed retailer websites (photos of products) | 37 | 6 |
| EN8 | Physical shop as well as online shop | 65 | 1 |
| EN9 | Trustworthy payment options other than credit cards | 45 | 4 |
| EN10 | Others | 7 | 10 |

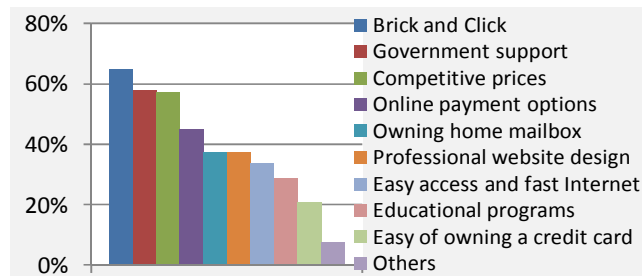

Figure 2. Factors facilitating online purchases by Saudi customers from Saudi e-retailers

The top two enablers of Saudi online purchases are both measures with the potential to alleviate the trust/experience issues identified above as key inhibitors: EN8 (the presence of a physical shop as well as an online shopfront) is selected by 65% of the respondents (ranked 1) and EN6 (government support, supervision and control of e-commerce) is selected by 58% (ranked 2). Similarly, EN5 (provision of educational programs) is seen by 29% of respondents as an enabler (ranked 8).

Other enablers that can be readily anticipated by the above analysis of inhibitors are EN9 (development of trustworthy and secure payment options other than credit cards) which is cited by 45% of the respondents (ranked 4), EN4 (local banks making it easier for customers to own credit cards) which is selected by 21% of the respondents (ranked 9), and EN1 (owning a home mailbox), chosen by 37 % (ranked 5).

A very important enabler (57%, ranked 3) that is not directly anticipated by the inhibitors analysis is EN3 (e-retailers being able to offer competitive prices). Similarly, EN7 (professional and easy-to-understand retailer websites, including product photos and complete specifications) is seen by 37% of the respondents as an enabler (ranked 6). Both of these results point to the type of features or attractions that Saudi e-retailers will need to provide to succeed.

A result that seems slightly anomalous, given other existing information about Internet access in Saudi Arabia, is that EN2 (easy access and fast Internet speed) is nominated as an enabler by a significant percentage of the respondents (33%, ranked 7).

Delving more deeply into the details of the individual responses allows us to gain further insights into key inhibitors of the demand for goods and services provided by the Saudi e-retail sector. Table 3 presents data relating to the interactions between inhibitors and customer attributes.

Table 3. Inhibitors of purchases from Saudi e-retailers and attributes of potential Saudi customers

| Inhibitors and Attributes of Customers | | | | | | | | | | | |
|---|---|---|---|---|---|---|---|---|---|---|---|
| | | Percentage of respondents selecting | | | | | | | | | |
| | % of sample | IN1 | IN2 | IN3 | IN4 | IN5 | IN6 | IN7 | IN8 | IN9 | IN10 |
| All respondent customers | 100.0 | 40.8 | 38.6 | 7.3 | 58.0 | 31.8 | 38.6 | 27.7 | 20.9 | 53.4 | 44.7 |
| Males | 50.0 | 33.0 | 43.2 | 7.8 | 51.0 | 32.5 | 41.3 | 28.2 | 16.5 | 64.1 | 46.6 |
| Females | 50.0 | 48.5 | 34.0 | 6.8 | 65.0 | 31.1 | 35.9 | 27.2 | 25.2 | 42.7 | 42.7 |
| Age 15-25 | 27.2 | 50.0 | 44.6 | 8.9 | 58.9 | 34.8 | 42.0 | 24.1 | 27.7 | 48.2 | 42.9 |
| Age 26-35 | 49.0 | 34.7 | 37.6 | 7.9 | 56.9 | 32.2 | 38.6 | 24.3 | 17.3 | 56.9 | 46.0 |
| Age 36 and over | 23.8 | 42.9 | 33.7 | 4.1 | 59.2 | 27.6 | 34.7 | 38.8 | 20.4 | 52.0 | 43.9 |
| Main city residents | 66.0 | 37.9 | 36.0 | 4.4 | 59.2 | 27.2 | 38.2 | 29.0 | 17.3 | 53.7 | 42.3 |
| Small city residents | 28.2 | 46.6 | 48.3 | 13.8 | 56.9 | 43.1 | 38.8 | 25.0 | 27.6 | 55.2 | 49.1 |
| Very small city residents | 5.8 | 45.8 | 20.8 | 8.3 | 50.0 | 29.2 | 41.7 | 25.0 | 29.2 | 41.7 | 50.0 |
| Have no mailbox | 49.5 | 51.5 | 34.3 | 6.9 | 55.9 | 40.7 | 35.3 | 23.5 | 22.5 | 47.1 | 45.1 |
| Mailbox at agency | 37.4 | 30.5 | 42.2 | 8.4 | 59.7 | 27.9 | 40.9 | 29.2 | 18.8 | 64.3 | 45.5 |
| House mailbox | 13.1 | 29.6 | 44.4 | 5.6 | 61.1 | 9.3 | 44.4 | 38.9 | 20.4 | 46.3 | 40.7 |
| Have credit card | 42.0 | 23.1 | 43.9 | 6.4 | 53.8 | 28.3 | 42.2 | 31.2 | 12.7 | 66.5 | 49.7 |
| No credit card | 57.3 | 54.2 | 35.2 | 8.1 | 61.9 | 34.7 | 36.4 | 25.4 | 27.1 | 44.5 | 41.5 |
| Home access to internet | 90.3 | 40.9 | 38.7 | 3.5 | 58.6 | 30.4 | 38.2 | 27.2 | 21.0 | 54.6 | 44.4 |
| No home access to Internet | 9.7 | 40.0 | 37.5 | 42.5 | 52.5 | 45.0 | 42.5 | 32.5 | 20.0 | 42.5 | 47.5 |
| Have bought online | 41.5 | 15.2 | 45.0 | 7.0 | 48.5 | 25.1 | 43.9 | 23.4 | 12.9 | 63.2 | 45.6 |
| Have not bought online | 58.5 | 58.9 | 34.0 | 7.5 | 64.7 | 36.5 | 34.9 | 29.9 | 26.6 | 46.5 | 44.0 |
| Bought online from KSA | 14.1 | 24.1 | 36.2 | 3.4 | 56.9 | 19.0 | 39.7 | 27.6 | 22.4 | 55.2 | 41.4 |
| Bought from foreigners only | 27.4 | 10.6 | 49.6 | 8.8 | 44.2 | 28.3 | 46.0 | 23.0 | 8.0 | 67.3 | 47.8 |

Data presented in Table 3 suggest that despite recent efforts to encourage Saudi residents to acquire mailboxes, many still have no access to mailboxes (49.5% of our sample) and only a small minority have home mailboxes (13.1%). Inevitably this would have a negative influence on the feasibility of conducting e-retail purchases. Another interesting statistic is that more than one-half (57.3%) of the respondents don't have a credit card. To some extent this is related to a cultural issue over the question of credit and interest rates, and points to the desirability for alternative online payment arrangements that are secure and effective.

Our analysis thus far has been based on the entire sample of respondents. Additional useful insights can be gained by focusing on a subset of this sample, namely the respondents who have already made some online purchases in the past. As can be seen from Table 3, 41.5% of the participants report having done so. There is a strong association between owning a credit card and having bought online: Of the 173 respondents who own a credit card, 116 say that they have bought online, a much higher percentage (67.05%) than for the sample as a whole.

Of the 171 respondents who have made online purchases, a majority have bought from overseas e-retailers only: they represent 27.4% of the whole sample, compared with 14.1% for those who have bought from vendors in Saudi Arabia (most, if not all, of the latter buyers have also bought from overseas vendors). From interviews for the previous qualitative analysis and from follow-up enquiries as part of the current analysis, it would appear that this (approximate) 2:1 imbalance is attributable largely to the reputation of global online retailers such as Amazon.com, eBay, Dell, etc. and the major airlines and hotel operators. Positive recommendations from relatives and friends tend to reinforce this pattern. In view of the above discussion about trust issues, it is relevant to note that e-retailers that could follow eBay's example and use customer feedback to substantiate the seller's trustworthiness would be able to alleviate Saudi buyers' customary lack of trust. Similarly, the use of PayPal for online payments can help build customer trust, because it acts as a third party between the seller and the buyer, thus helping to resolve problems that might arise.

Examination of the bottom rows of Table 3 reveals interesting contrasts and similarities between respondents with and without online purchase experiences. For example, it can be seen clearly that whilst IN1 (lack of experience in buying online) is a major inhibitor to respondents who have not bought online (it is selected by 59.0% of them, ranked 2) it is of little relevance to respondents who have done so, especially to those who have bought from overseas vendors only (10.6%). Yet as far as the issues of trust in Saudi e-retailers (IN2, IN4, IN6, and IN10) and the missing presence of government regulations and supervision (IN9) the various categories of respondents give very similar ratings. These similarities serve to reinforce a general conclusion that has emerged from our analysis: trust in Saudi e-retailers (or the lack of it) is probably the most important factor affecting current and potential Saudi customers. From this, it is reasonable to draw the implication that the government can play, if it so wishes, a key role in regulating, supervising, and facilitating e-retail in Saudi Arabia. The critical question, then, becomes whether there are valid justifications for the government to take such an interventionist role in normal commerce, as opposed to the cases of e-government and e-learning which involve public services or "social" goods. Such a question must be left to future research.

## 5. CONCLUSION

This paper has investigated the issues that positively and negatively influence customers to purchase from online retailers in Saudi Arabia. The study comes up with a list of factors that influence the decision of Saudi customers to purchase from online retailers in Saudi Arabia ranked according their rating. The main finding of this study clearly demonstrates that the most serious inhibitor tend to be related to a lack of trust. It also appears that in the minds of many Saudi customers, a lack of clear government regulations and legislations on e-commerce may have been a key contributor to this general lack of trust and experience (chosen by 53.4% of respondents, ranked 2). On the other hand, the top two enablers of Saudi online purchases are both measures with the potential to alleviate the trust/experience issues identified above as key inhibitors: EN8 (the presence of a physical shop as well as an online shopfront) is selected by 65% of the respondents (ranked 1) and EN6 (government support, supervision and control of e-commerce) is selected by 58% (ranked 2).

In conclusion, trust in Saudi e-retailers (or the lack of it) is probably the most important factor affecting current and potential Saudi customers. From this, it is reasonable to draw the implication that the government can play, if it so wishes, a key role in regulating, supervising, and facilitating e-retail in Saudi Arabia. The critical question, then, becomes whether there are valid justifications for the government to take such an interventionist role in normal commerce, as opposed to the cases of e-government and e-learning which involve public services or "social" goods. Such a question must be left to future research.

This study is limited in terms of the sample size. However, the data collection still ongoing and by the end of the study the sample is expected to reach 700 participants. This study is still in progress. We will, in due course, be able to report all the factors that positively and negatively affect e-retailing growth in Saudi Arabia and gaining the information from all involved parties (i.e. retailers, customers and government) in this field in order to contribute to e-commerce development in general and the diffusion of online retailing in particular in Saudi Arabia.